\documentclass[aps,prd,superscriptaddress,nofootinbib,eqsecnum,twocolumn,preprintnumbers]{revtex4-1}

\pdfoutput=1

\usepackage{latexsym}
\usepackage{amsfonts}
\usepackage{amsmath}
\usepackage{amssymb}
\usepackage{graphicx,color}
\usepackage{float}
\usepackage{hyperref}
\usepackage{subfigure}
\usepackage{dcolumn}% Align table columns on decimal point
\usepackage{soul}
\usepackage{ulem}
\usepackage{verbatim}
\usepackage{xcolor}
\begin{document}

\title{Viability of warm inflation with standard model interactions}

\author{Rudnei O. Ramos} 
\email{rudnei@uerj.br}
\affiliation{Departamento de Fisica Teorica, Universidade do Estado do
  Rio de Janeiro, 20550-013 Rio de Janeiro, RJ, Brazil }

\author{Gabriel S. Rodrigues} 
\email{gabriel.desenhista.gr.gr@gmail.com}
\affiliation{Departamento de Fisica Teorica, Universidade do Estado do
  Rio de Janeiro, 20550-013 Rio de Janeiro, RJ, Brazil }

%%%%%%%%%%%%%%%%%%%%%%%%%%%%%%%%%%%%%%%%%%%%%%%%%
\begin{abstract}

The minimal warm inflation scenario proposed in Ref.~\cite{Berghaus:2025dqi} --- featuring an 
axionlike inflaton coupled to Standard Model (SM) gluons via the standard interaction $\phi G \tilde G$ --- offers 
a compelling bridge between inflationary dynamics and SM particle content. 
While the model retains only the inflaton as a beyond-SM field, 
its original analysis relied on some approximate treatments of warm inflation's (WI) dynamics. 
Here, we revisit this scenario using \texttt{WI2easy}, a precision computational tool for WI dynamics~\cite{Rodrigues:2025neh},
to rigorously evaluate the model's viability and full range of model's parameters compatible with the observational 
parameters. Overall, we find that the results of Ref.~\cite{Berghaus:2025dqi} hold, but with significant
differences in the weak and strong dissipative regimes of WI.

\end{abstract}

\maketitle

%%%%%%%%%%%%%%%%%%%%%%%%%%%%%%%%%%%%%%%%%%%%%%%
\section{Introduction}

Warm inflation (WI) is a cosmological framework where the inflaton field interacts continuously with a thermal 
radiation bath during the accelerated expansion phase, generating entropy (particles) through dissipative dynamics 
rather than requiring a separate reheating epoch~\cite{Berera:1995ie,Berera:1996nv,Berera:1998gx,Berera:1996fm}. 
Unlike cold inflation (CI), which assumes a near-zero temperature universe until reheating begins, WI maintains a 
thermalized environment via particle interactions, with dissipation encoded in a coefficient $\Upsilon$, which typically can
be a function of the temperature and the inflaton amplitude, besides of microscopic parameters entailing the microscopic 
derivation of it. The early ground breaking work on WI has shown how thermal fluctuations---rather than quantum 
ones---could 
seed primordial density perturbations, while key developments included deriving consistent perturbation equations for 
dissipative systems~\cite{Graham:2009bf,Bastero-Gil:2011rva,Ramos:2013nsa,Bastero-Gil:2014jsa} and calculating modified 
observational signatures. In order to preserve the flatness of the inflaton potential and prevent it from receiving 
large thermal corrections which could disrupt the inflationary attractor solution, early models of WI explored 
SUSY-inspired interactions (for a review, see e.g. Ref.~\cite{Berera:2008ar}), 
while later studies tried to better integrate WI with particle physics frameworks making use of interactions where the inflaton would be a
pseudo-Goldstone boson~\cite{Bastero-Gil:2016qru,Bastero-Gil:2019gao} or to have axionlike couplings to gauge fields~\cite{Berghaus:2019whh}, 
such as to protect the inflaton potential from large quantum and thermal corrections
due to the models' symmetries\footnote{{}For earlier WI models based on axion inflation, see also Refs.~\cite{Mishra:2011vh,Visinelli:2011jy,Kamali:2019ppi}.}  
(for a recent
review on model building proposals for WI, see. e.g.~Ref.~\cite{Kamali:2023lzq}).

Recently, the authors of Ref.~\cite{Berghaus:2025dqi} have proposed a WI construction with a minimal extension of the standard model (SM) of
particle physics, where the inflaton is coupled to gluons via the standard $\phi G \tilde G$ axionlike 
interaction. Some earlier attempts of building WI and WI motivated models
with SM interactions include for example Refs.~\cite{Dymnikova:2000gnk,Dymnikova:2001jy,Kamada:2009hy,Levy:2020zfo}.
In particular, in the Ref.~\cite{Kamada:2009hy}, the authors considered the  minimal
supersymmetric standard model, with the inflaton considered to be a flat direction in the model. The model was, however, unsuccessful for WI because of the presence of large thermal corrections to the inflaton potential~\footnote{ Note, however, that one of the first
working models for WI was based on supersymmetry and proposed in Ref.~\cite{Berera:2002sp}, with dissipation coefficients originally derived
in Ref.~\cite{Moss:2006gt}.}.  The work of Ref.~\cite{Levy:2020zfo} implemented a SM WI extension with a neutrino portal and using the symmetries and construction
of the warm little inflaton model
proposed in the Refs.~\cite{Bastero-Gil:2016qru,Bastero-Gil:2019gao}.
The possibility of having a WI model built with genuine SM interactions is a major result. 
Constructing a WI model within the SM framework offers compelling theoretical and phenomenological advantages. 
By leveraging known SM interactions and minimal extensions of them---such as the inflaton coupling to the gluons via 
the axionlike interaction---this model not only minimizes beyond-the-SM (BSM) particle content, but also enhances the 
testability of the WI general predictions and reduces possible fine-tuning in previous WI constructions. 
In this context, the thermal bath in WI naturally arises from SM particle production, sidestepping ad hoc 
reheating mechanisms, while providing a seamless transition to the radiation-dominated era. 
Crucially, the SM-derived dissipation coefficient comes from sphaleron decay, which is  a most studied subject in the context of
gauge field thermodynamics~\cite{Moore:2010jd,Arnold:1999ux,Arnold:1999uy,Laine:2022ytc,Drewes:2023khq}.
{}Furthermore, by having WI embed with SM interactions enables a formulation grounded in particle physics experimentally, 
enabling cross-validation through collider constraints or early universe thermodynamics. 

We revisit the WI scenario with SM interactions proposed in Ref.~\cite{Berghaus:2025dqi} using the \texttt{WI2easy} 
code~\cite{Rodrigues:2025neh}, which replaces the explicit stochastic perturbation 
methods commonly used previously in WI~\cite{Graham:2009bf,Bastero-Gil:2011rva,Bastero-Gil:2014jsa,Montefalcone:2023pvh} 
with a deterministic Fokker-Planck formalism~\cite{Ballesteros:2022hjk,Ballesteros:2023dno}. 
While mathematically equivalent to prior approaches, the {}Fokker-Planck framework offers two key advantages:
simplified numerical implementation and enhanced precision.
\texttt{WI2easy} is then a precision computational tool for WI dynamics. As such, we
can rigorously evaluate the model's viability, bypassing approximations used in the analysis done in Ref.~\cite{Berghaus:2025dqi}.
In particular, we quantify deviations from Ref.~\cite{Berghaus:2025dqi} approximations, revealing some significant differences in both
weak and strong dissipation regimes of WI, although in intermediate regimes the results of that reference holds approximately well.

The organization of the paper is as follows.
In Sec.~\ref{sec2}, we briefly introduce the construction of Ref.~\cite{Berghaus:2025dqi}. We also
take advantage of some of the well-known thermodynamics and kinetic theory for quantum chromodynamics (QCD)
to explore advances in the study considered in here.
In Sec.~\ref{sec3}, we apply the \texttt{WI2easy}  to the problem under study and give the main results
obtained from our analysis.
In Sec.~\ref{conclusions}, we give our conclusions and final remarks.

%%%%%%%%%%%%%%%%%%%%%%%%%%%%%%%%%%%%%%%%%%%%%%%%%%%%%%%%%%%%%%%
\section{WI with SM interactions}
\label{sec2}

WI is characterized by the evolution equations for the inflaton 
field $\phi$ and radiation energy density $\rho_r$,
\begin{eqnarray}
&&\ddot \phi + (3 H + \Upsilon) \dot \phi + V_{,\phi}=0,
\label{eqphi}
\\
&&\dot \rho_r + 4 H \rho_r = \Upsilon \dot \phi^2,
\label{eqrhor}
\end{eqnarray}
where $\Upsilon$ is the dissipation coefficient, which is assumed to be derived from first principles
quantum field theory (for earlier explicit  derivations, see for example
Refs.~\cite{Gleiser:1993ea,Berera:1998gx,BasteroGil:2010pb,Bastero-Gil:2016qru,Bastero-Gil:2019gao}). 
{}For a thermalized radiation bath, it also follows that 
 \begin{eqnarray}
\rho_{r}=\frac{g_* \pi^2}{30} T^4,
\label{rhor}
 \end{eqnarray}
where $g_*$ is the radiation bath
degrees of freedom. {}For the temperatures that we obtain here, $T \gtrsim 10^{11}$ GeV,  we can
assume that all SM particles are thermalized,
$g_* = 106.75$, while when also considering the inflaton field thermalized,
then $g_* = 106.75 + 1$.

In the WI construction of Ref.~\cite{Berghaus:2025dqi}, it is made use of an
axionlike  field making
the role of the inflaton and which is coupled to the QCD gluon through the dimension five interaction, 
\begin{eqnarray}
{\cal L}_{\rm int} &=& - \frac{\alpha_g}{8 \pi f_a}\phi  G_{\mu \nu}^c \tilde{G}^{c\,
    \mu \nu},
\label{Lint}
\end{eqnarray}
where $c=1,\ldots, N_c^2-1$, $N_c=3$, $G_{\mu \nu}^c$ is the gluon field tensor,
$\tilde{G}^{c\,\mu \nu}$ is its dual and $\alpha_g = g^2/(4\pi)$, with $g$ the strong coupling
gauge constant and $f_a$ is the axion decay constant.

As first realized in Ref.~\cite{Berghaus:2019whh}, the dimension five axionlike interaction
to $SU(N_c)$ gauge fields can lead to a successful WI realization where the dissipation 
coefficient $\Upsilon$ is a consequence of sphaleron decay in a thermal bath. 
In a $SU(N_c)$ gauge field theory,
the sphaleron rate is~\cite{Moore:2010jd}
\begin{equation}
\Gamma_{\rm sphal} \simeq  \kappa \left(\alpha_g N_c\right)^5 T^4,
\label{Gsphal}
\end{equation}
where the constant $\kappa$ here is given by
\begin{equation}
\kappa \simeq 0.21 \left( \frac{N_c g^2 T^2}{m_D^2} \right) 
\frac{N_c^2-1}{N_c^2}\left[\ln\left(\frac{m_D}{\gamma}\right)+3.041\right],
\label{kappa}
\end{equation}
where $m_D$ is color Debye mass,
\begin{equation}
m_D^2 = \left( 2 N_c^2 + N_f\right) \frac{g^2 T^2}{6},
\label{mD}
\end{equation}
with $N_f$ the number of quark flavors and $\gamma$ is the rate of color randomization,
\begin{equation}
\gamma=\frac{g^2 N_c T}{4 \pi}
\left[\ln\left(\frac{m_D}{\gamma}\right)+3.041\right].
\end{equation}
Explicitly, $\kappa$ is found to be given by 
\begin{eqnarray}
\kappa & = & 1.26 \frac{N_c^2-1}{(2 N_c+N_f) N_c} W\left(
\frac{e^{3.041}}{N_c}\sqrt{\frac{2\pi(2N_c+N_f)}{3 \alpha_g}}\right),
\nonumber \\
\label{kappa2}
\end{eqnarray}
where $W(x)$ is the
Lambert function, given by the principal solution of $x=w e^w$. 
Here, we also consider the running of the gluon gauge 
coupling as a function of the temperature, $\alpha_g\equiv \alpha_g(T)$, 
which at one-loop order is~\cite{Graf:2010tv}
\begin{eqnarray}
\alpha_g^{-1}(T) &=& \alpha_g^{-1}(M_Z) + \frac{11N_c - 2 N_f}{6 \pi} \ln \left(\frac{T}{M_Z} \right),
\nonumber \\
\label{alphagT}
\end{eqnarray}
where $\alpha_g(M_Z) = 0.1172$ with $M_Z=91.188$GeV.

%%%%%%%%%%%%%%%%%%%%%%%%%%%%%%%%%%%%%%%%%%
\begin{center}
\begin{figure}[!bth]
\subfigure[]{\includegraphics[width=7.5cm]{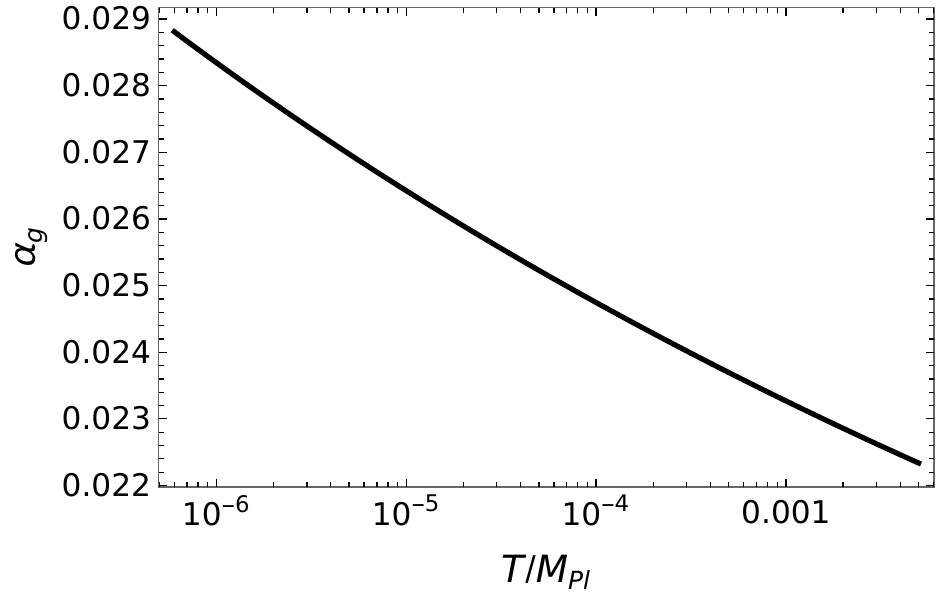}}
\subfigure[]{\includegraphics[width=7.5cm]{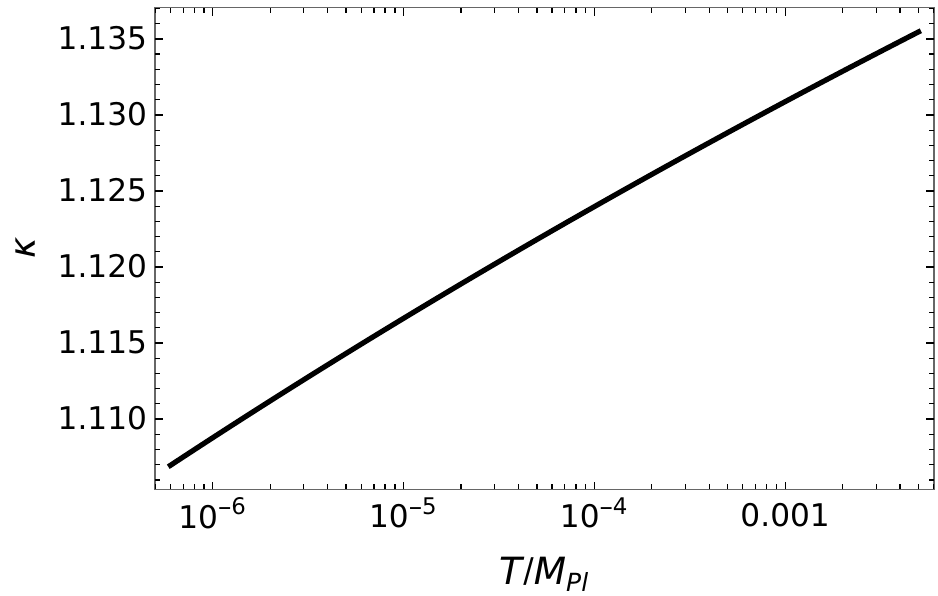}}
\caption{The temperature dependence of $\alpha_g$ (panel a) from Eq.~(\ref{alphagT})
and the coefficient $\kappa$ (panel b) in the sphaleron rate Eq.~(\ref{Gsphal}).}
\label{fig1}
\end{figure}
\end{center}
%%%%%%%%%%%%%%%%%%%%%%%%%%%%%%%%%%%%%%%%%%%  

As we are going to verify in the next section, in the observational region of interest for the SM WI model,
the typical  temperatures that we find have values that 
fall in the range $T \in [6\times 10^{-7},5\times 10^{-3}] M_{\rm Pl}$,
where $M_{\rm Pl}$ is the reduced Planck mass, $M_{\rm Pl} \simeq 2.44\times 10^{18}$ GeV.
Using Eqs.~(\ref{kappa2}) with (\ref{alphagT}), we show in {}Fig.~\ref{fig1} the overall temperature dependence
of the strong fine structure coupling $\alpha_g$ and also the dimensionless factor $\kappa$ in the temperature range of interest.
In particular, the result for $\kappa$ is consistent with the value $\kappa \sim 1$ assumed in Ref.~\cite{Berghaus:2025dqi}, but $\kappa$ and also $\alpha_g$
still display some relevant temperature dependence. Hence, we keep in our study the full $\kappa(T)$ result
from Eq.~(\ref{kappa2}), considering also Eq.~(\ref{alphagT}).

{}Finally, the sphaleron rate $\Gamma_{\rm sphal}$ is related to the dissipation coefficient $\Upsilon$ in WI
as~\cite{Berghaus:2019whh,Laine:2021ego} 
\begin{equation}
\Upsilon_{\rm sph}= \frac{\Gamma_{\rm sphal}}{2 T f_a^2}.
\label{Upsilon}
\end{equation}
The result Eq.~(\ref{Upsilon}) applies in the pure gauge model in the absence of fermions. 
The presence of fermions (quarks) tends to suppress the sphaleron dissipation in the case of 
chiral symmetry breaking ($m_q\neq 0$) and can even make it completely vanish for
massless fermions~\cite{Berghaus:2020ekh,Drewes:2023khq,Berghaus:2024zfg}. The effect of the
expansion also changes the sphaleron dissipation expression, in which case the
effective dissipation in the SM WI model becomes~\cite{Berghaus:2025dqi}
\begin{equation}
\Upsilon_{\rm eff} = 
\frac{ \Upsilon_{\rm sph}}{1+ \frac{4 \tilde{N}_f f_a^2  \Upsilon_{\rm sph}}{H N_cT^2} } ,
\label{Upseff}
\end{equation}
where $\tilde{N}_f=5$ is the effective number of quarks contributing to $\Upsilon_{\rm eff}$, as considered 
and explained in 
Ref.~\cite{Berghaus:2025dqi}~\footnote{
Note that ultraviolet completions of axionlike models involve beyond SM heavy fields like,
for example, in the Kim-Shifman-Vainshtein-Zakharov (KSVZ) axion model~\cite{Kim:1979if,Shifman:1979if}, 
which introduces heavy exotic quarks and generates the axion coupling Eq.~(\ref{Lint}) through the chiral anomaly.
Even though the heavy exotic quarks decouples in the effective theory, they could still affect the dissipation
through extra decay channels much similar to the cases reported in Refs.~\cite{Berghaus:2020ekh,Drewes:2023khq,Berghaus:2024zfg}.
However, we can always choose appropriate parameters (e.g. the Yukawa couplings in these models) such that
the chiral decay rates from the interactions to remain much smaller that the Hubble rate, $\Gamma_{\rm ch} \ll H$ and
the result Eq.~(\ref{Upseff}) still remains valid.}. The Hubble parameter in Eq.~(\ref{Upseff}) is well approximated by the slow-roll result,
\begin{equation}
H \simeq \sqrt{ \frac{V(\phi)}{3 M_{\rm Pl}^2}}.
\label{Hsr}
\end{equation}
The expression Eq.~(\ref{Upseff}) turns out to be a complicate function of
both temperature and inflaton amplitude. As is known in WI~\cite{Graham:2009bf,Bastero-Gil:2011rva},
the functional dependence of the dissipation coefficient on both $T$ and $\phi$ can strongly
affect the perturbation spectrum and observable quantities derived from it. This effect
is traditionally modeled through a function $G(Q)$ of the dissipation ratio
$Q=\Upsilon/(3H)$ and that multiplies the scalar of curvature power spectrum (evaluated at Hubble exit)
in WI~\cite{Kamali:2023lzq},
\begin{eqnarray}
P_{\mathcal{R}}&=&\left(\frac{H^2}{2 \pi \dot\phi}\right)^2
 \left(1+2n_{*} + \frac{2\sqrt{3}\pi
  Q}{\sqrt{3+4\pi Q}}{T\over H}\right) G(Q)\Bigr|_{k=aH},
\nonumber \\
\label{powers}
\end{eqnarray}
where $n_*$ in the above expression denotes the possible statistical distribution for the inflaton
due to the presence of the radiation bath. The function $G(Q)$ can only be derived numerically,
by solving the complete set of perturbation equations in WI. Recent public codes have been
dedicated to the precise numerical determination of $G(Q)$ depending on the WI model used~\cite{Montefalcone:2023pvh,Rodrigues:2025neh}.
In Ref.~\cite{Berghaus:2025dqi}, the authors made use of a fitting form for the power spectrum in
WI proposed in the Ref.~\cite{Mirbabayi:2022cbt}, where
\begin{eqnarray}
P_{\mathcal{R}}\simeq \frac{H^3 T}{4 \pi^2 \dot\phi^2}F(Q)\Bigr|_{k=aH},
\label{powersFQ}
\end{eqnarray}
and with the function $F(Q)$ given by
\begin{eqnarray}
F(Q) &=& 3.27 \times 10^{-4} Q^7 
\nonumber \\
&+& 168 Q \left[ \frac{1}{3} \left( 1 + \frac{9Q^2}{25} \right) + \frac{2}{3}
\tanh \frac{1}{30Q} \right].
\nonumber \\
\label{FQ}
\end{eqnarray} 
Essentially, the authors of Ref.~\cite{Mirbabayi:2022cbt} have estimated the result Eq.~(\ref{powersFQ})
by considering two critical assumptions. {}First, that the dissipation coefficient was the one
coming from the sphaleron decay for pure gauge fields, Eq.~(\ref{Upsilon}), i.e., $\Upsilon \propto T^3$.
As already noticed from Eq.~(\ref{Upseff}), the effective dissipation coefficient 
$\Upsilon_{\rm eff}$ in the SM WI model
is expected to have some quite different functional form.
Second, they have neglected the statistical distribution term $n_*$ in Eq.~(\ref{powers}),
which means in particular that the inflaton perturbations should remain in the vacuum state throughout
the dynamics and never thermalizes.  Here we will test explicitly the validity of these two
hypothesis. 

Infer the thermalization of the inflaton perturbations in WI is a difficult problem. This is because
we need to know precisely the interactions of the inflaton field with all other degrees of freedom
to properly estimate the scattering rate of the inflaton field.
Then, we need to solve the kinetic equation for the inflaton particle distribution in the expanding spacetime to check whether
the statistical distribution for the inflaton field would indeed approach a thermal one.
In the Ref.~\cite{Bastero-Gil:2017yzb} it was shown that in WI, even for scattering rates much smaller than the
Hubble rate, the inflaton distribution can acquire a shape close to a thermal one. Then when
$\Gamma_{\rm scat} \gtrsim H$, the distribution approaches a thermal equilibrium form, given
by a Bose-Einstein distribution, $n_* \to n_{\rm BE}$, and as indeed expected in general.
Here, we can take advantage of the existent literature on the thermodynamics and kinetic theory of
axions in QCD to then infer the possibility of thermalization of the inflaton field.
Despite the fact that there is still a debate in the literature about the thermalization
of axions in QCD~\cite{Salvio:2013iaa,Bouzoud:2024bom}, the temperatures we are interest
in are sufficiently high that we can make use of the hard thermal loop estimates of Ref.~\cite{Masso:2002np}.
The dominant process for axion thermalization comes from the scattering processes with the gluon,
$\phi+g \leftrightarrows g + g$, while the processes involving quarks are subdominant.
Summing all processes $\Gamma_{\rm scat}=\Gamma_{\phi + g \leftrightarrows g +g} +
\Gamma_{\phi + q \leftrightarrows g +q}+
\Gamma_{\phi + \bar q \leftrightarrows g +\bar q} + \Gamma_{\phi + g \leftrightarrows q +\bar q}$, the result is~\cite{Masso:2002np}
\begin{eqnarray}
\Gamma_{\rm scat}\!\!\!&=& \!\!\!\frac{\alpha_g^3 T^3}{f_a^2} \left\{ \frac{15 \zeta(3)}{2 \pi^4}
\left[ 3 \!-\! \gamma_E \!+\! \frac{2 \zeta'(3)}{\zeta(3)} \!-\! \ln (2 \pi \alpha_g)
\!-\! \frac{17}{12} \right] \right.
\nonumber \\
&+& \left.  
\frac{N_f\zeta(3)}{6 \pi^4} \right. 
\nonumber \\
&+&\left.  \frac{N_f\zeta(3)}{2 \pi^4}
\left[ \frac{1}{4} \left( 9 + \ln 4 - 6 \gamma_E + \frac{6 \zeta'(3)}{\zeta(3)} \right)
\right.\right.
\nonumber \\
&+& \left.\left. \frac{3}{4} \left(3 - 2 \gamma_E +  \frac{2 \zeta'(3)}{\zeta(3)} \right)
-\frac{3}{2} \ln(2 \pi \alpha_g) - \frac{15}{8} \right]
 \right\},
\nonumber \\
\label{Gammascat}
\end{eqnarray}
where $\zeta(x)$ is the zeta-function and $\gamma_E$ is the Euler-Mascheroni constant, 
$\gamma_E \simeq 0.577$.

The result given by Eq.~(\ref{Gammascat}) will be directly compared with the Hubble parameter
in the next section. {}Finally, the suitability of the approximated power spectrum, Eq.~(\ref{powersFQ}), 
is compared to the one obtained from the full numerical solution of perturbations in WI using \texttt{WI2easy}.

%%%%%%%%%%%%%%%%%%%%%%%%%%%%%%%%%%%%%%%%%%
\section{Background dynamics and perturbation quantities in the SM WI model}
\label{sec3}

In our analysis of the SM WI model we will make use of the recently released code
\texttt{WI2easy} for WI analysis of background and perturbation dynamics.
The code can handle general forms of dissipation coefficients, 
$\Upsilon = C_\Upsilon f(T,\phi)$, which makes it well-suited to treat to effective dissipation
coefficient Eq.~(\ref{Upseff}). Here, we identify the dimensionless constant $C_\Upsilon$ 
with the axion constant through
\begin{equation}
C_\Upsilon = \frac{M_{\rm Pl}^2}{f_a^2}.
\label{CUps}
\end{equation}
\texttt{WI2easy} automatically determines the numerical value for $C_\Upsilon$ once an initial value for
the dissipation ratio $Q$ is given, while also performing the normalization
of the inflaton potential 
such that at the pivot scale the power spectrum has the amplitude
according to the CMB normalization. Note that the dissipation ratio is here defined
as 
\begin{equation}
Q = \frac{\Upsilon_{\rm eff}}{3H}.
\label{Q}
\end{equation}
The pivot scale has the value $k_p=0.05 {\rm Mpc}^{-1}$, while  
the CMB amplitude normalization is $A_s \simeq 2.105 \times 10^{-9}$ (we follow 
Planck's default values~\cite{Aghanim:2018eyx}). {}For the inflaton potential,
we consider the same used in Ref.~\cite{Berghaus:2025dqi}, such that our results 
can be directly compared to,
\begin{equation}
V(\phi) = \lambda \phi^4 \equiv \frac{V_0}{M_{\rm Pl}^4} \phi^4,
\label{Vphi}
\end{equation}
where the potential normalization $V_0$ is automatically computed in \texttt{WI2easy}
such that the the spectrum will have the proper normalization $A_s$ at Hubble exit.

Under the assumptions used by the authors of Ref.~\cite{Mirbabayi:2022cbt} to
obtain Eq.~(\ref{powersFQ}), the equivalent in the \texttt{WI2easy} options available in its
built-in interface is to
choose the option of including explicitly a radiation noise term in the
equation for the radiation perturbation (see the discussion in Ref.~\cite{Bastero-Gil:2014jsa}
where this term was first identified and also the \texttt{WI2easy} paper~\cite{Rodrigues:2025neh}
for additional discussions in connection to the code results). As already
mentioned, Ref.~\cite{Mirbabayi:2022cbt} has also implicitly assumed $n_*=0$
in Eq.~(\ref{powers}). In fact they have dropped the entire $1+ 2 n_*$ term in Eq.~(\ref{powers}),
which makes their expression for the power spectrum suitable only for $Q\gtrsim 1$, 
i.e., in the strong dissipative regime of WI. Hence, comparing Eq.~(\ref{powers})
with (\ref{powersFQ}), we can identify their approximated $G(Q)$ function as being
\begin{equation}
G_{\rm approx}(Q) = \frac{\sqrt{3+4 \pi Q}}{2 \sqrt{3} \pi Q} F(Q).
\label{Gapprox}
\end{equation}
We compare Eq.~(\ref{Gapprox}) with the results provided by \texttt{WI2easy}.

%%%%%%%%%%%%%%%%%%%%%%%%%%%%%%%%%%%%%%%%%%
\begin{center}
\begin{figure}[!bth]
\includegraphics[width=7.5cm]{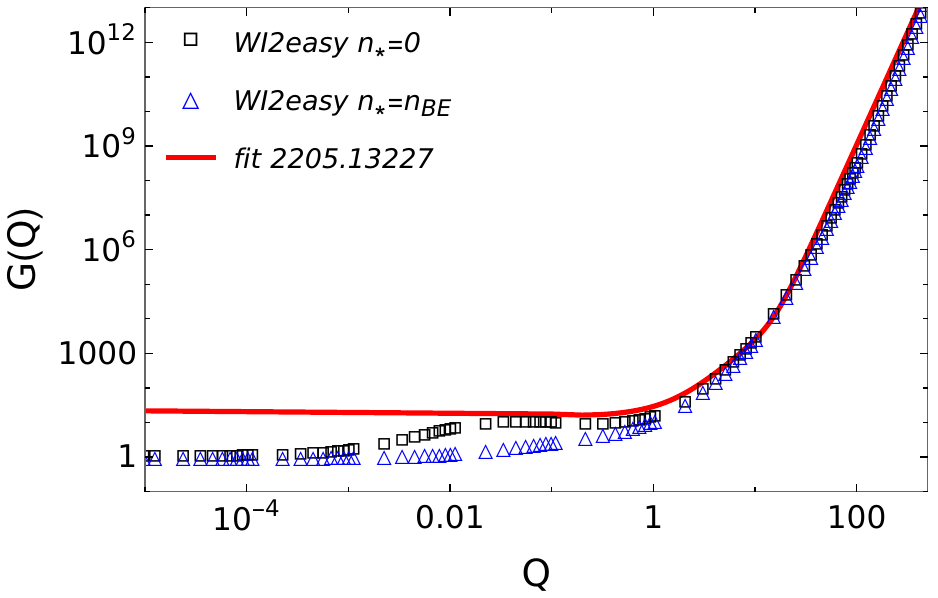}
\caption{The $G(Q)$ function from the fitting of Ref.~\cite{Mirbabayi:2022cbt}
compared with the results from \texttt{WI2easy} in the absence and presence of
inflaton thermalization.}
\label{fig2}
\end{figure}
\end{center}
%%%%%%%%%%%%%%%%%%%%%%%%%%%%%%%%%%%%%%%%%%%  

By fully implementing the SM WI model in \texttt{WI2easy}, we first make the
comparison of Eq.~(\ref{Gapprox}) with our numerical result. This is 
shown in {}Fig.~\ref{fig2}. Among the options available in \texttt{WI2easy}, we consider the
cases
of nonthermalized ($n_*=0$) and fully thermalized  ($n_*=n_{\rm BE}$)
inflaton perturbations. One notices that Eq.~(\ref{Gapprox}) only roughly fits
the numerical results, for both nonthermalized and thermalized
inflaton perturbations, for $Q\gtrsim 1$, while in the weak dissipative regime,
$Q \lesssim 1$, it displays noticeable discrepancies, in particular in the case of thermalized inflaton 
perturbations.

%%%%%%%%%%%%%%%%%%%%%%%%%%%%%%%%%%%%%%%%%%
\begin{center}
\begin{figure}[!bth]
\includegraphics[width=7.5cm]{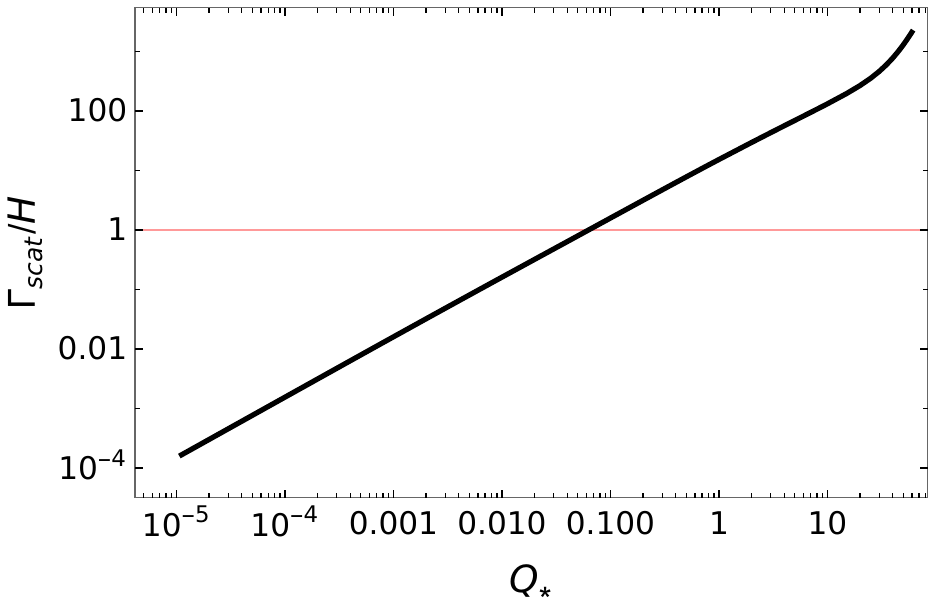}
\caption{The ratio of the scattering rate for the inflaton field, Eq.~(\ref{Gammascat}), with the Hubble parameter
as a function of $Q$ at Hubble exit.}
\label{fig3}
\end{figure}
\end{center}
%%%%%%%%%%%%%%%%%%%%%%%%%%%%%%%%%%%%%%%%%%%  

%%%%%%%%%%%%%%%%%%%%%%%%%%%%%%%%%%%%%%%%%%
\begin{center}
\begin{figure}[!bth]
\subfigure[]{\includegraphics[width=7.5cm]{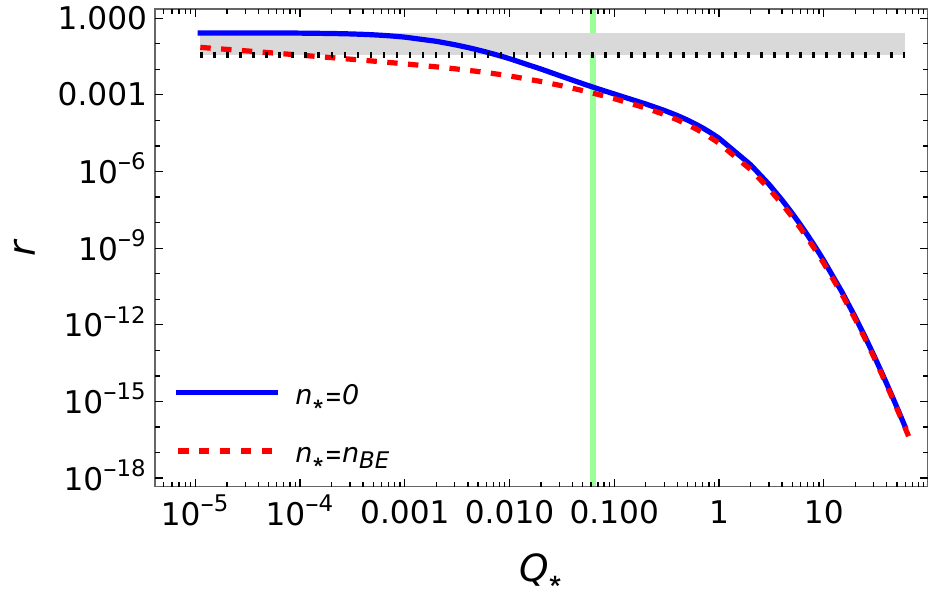}}
\subfigure[]{\includegraphics[width=7.5cm]{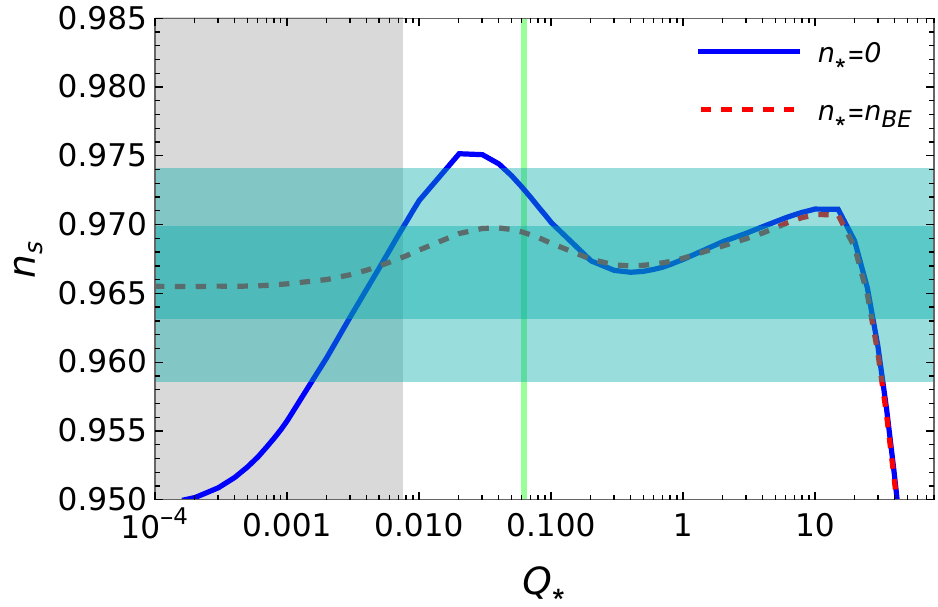}}
\subfigure[]{\includegraphics[width=7.5cm]{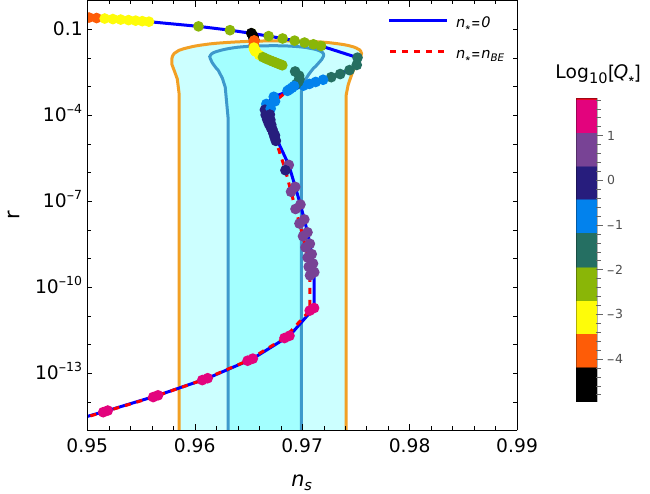}}
\caption{The tensor-to-scalar ratio (panel a) and the spectral tilt (panel b) at different values of $Q_*$. 
The tensor-to-scalar as a function of $n_s$ is presented in panel c. The top gray region in panel a
and the left gray region in panel b indicate the region where $r> 0.036$, with the upper bound on $r$ from the
BICEP, Keck Array and Planck combined data~\cite{BICEP:2021xfz}. The cyan regions in panels b
and c indicate the one- and two-sigma constrains also from Ref.~\cite{BICEP:2021xfz}. The vertical green line
in panels a and b indicates the expected threshold for thermalization ($n_*=n_{\rm BE}$).}
\label{fig4}
\end{figure}
\end{center}
%%%%%%%%%%%%%%%%%%%%%%%%%%%%%%%%%%%%%%%%%%%
To check for thermalization of the inflaton perturbations, we run \texttt{WI2easy}
for a range of values of $Q$ at Hubble exit time. The temperature is obtained at the different values
of $Q$ and used to check the ratio of the scattering rate Eq.~(\ref{Gammascat})
with the Hubble parameter, $\Gamma_{\rm scat}/H$, evaluated at the Hubble exit point $k=aH$. The results are shown
in {}Fig.~\ref{fig3}. The result of {}Fig.~\ref{fig3} indicates that the inflaton perturbations
are expected to thermalize for $Q_* \gtrsim 0.08$, hence, already in the weak dissipative regime of WI. 
This result, together with the ones shown in {}Fig.~\ref{fig2}, indicates that the power spectrum
can have significant differences in the thermal case ($n_*=n_{BE}$) than in the nonthermalized situation,
in special in the weak dissipative regime of WI.

%%%%%%%%%%%%%%%%%%%%%%%%%%%%%%%%%%%%%%%%%%
\begin{center}
\begin{figure*}[!bth]
\subfigure[]{\includegraphics[width=7.5cm]{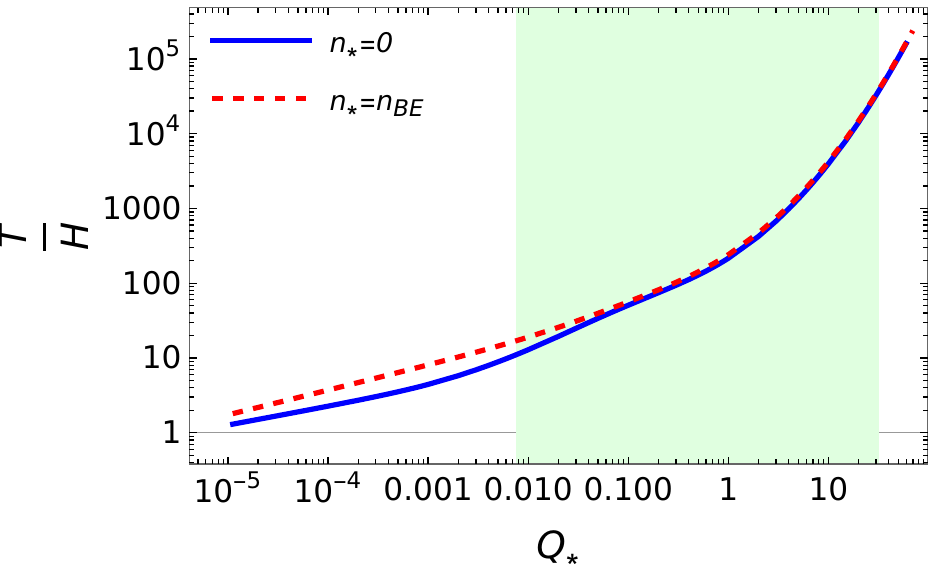}}
\subfigure[]{\includegraphics[width=7.5cm]{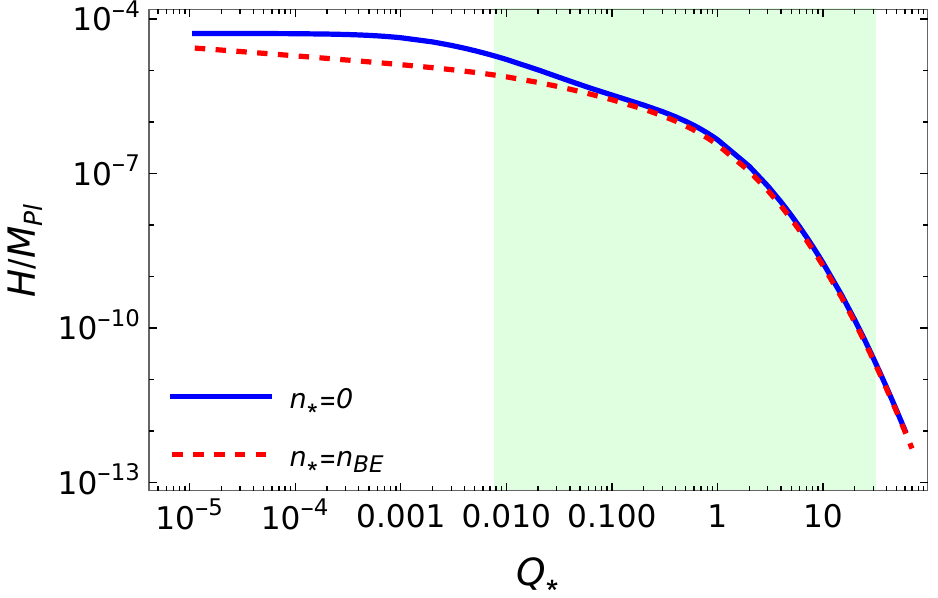}}
\subfigure[]{\includegraphics[width=7.5cm]{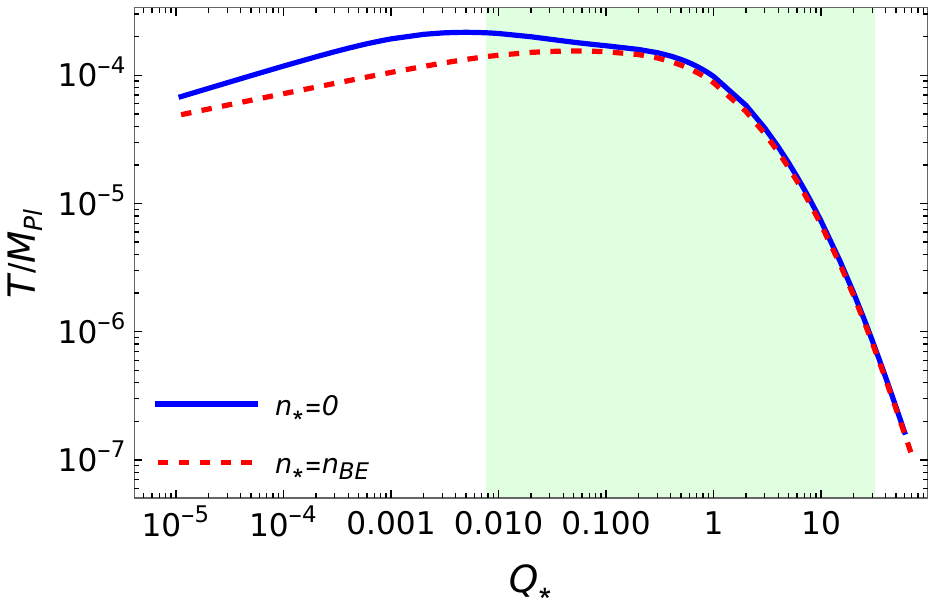}}
\subfigure[]{\includegraphics[width=7.5cm]{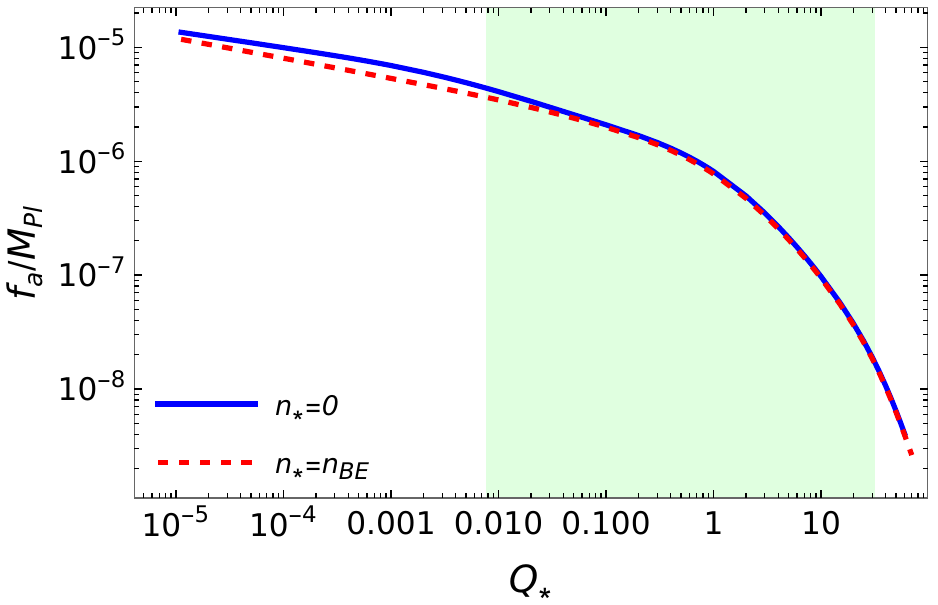}}
\caption{Some of the key background quantities in WI, namely $T/H$ (panel a), 
$H$ (panel b), $T$ (panel c) and the axion decay constant
$f_a$ (panel d). The light green band in all the plots indicates the region 
of $Q_*$ values for which $r$ and $n_s$ are within the observational 
window (see {}Fig.~\ref{fig4}).}
\label{fig5}
\end{figure*}
\end{center}
%%%%%%%%%%%%%%%%%%%%%%%%%%%%%%%%%%%%%%%%%%%

Still using  \texttt{WI2easy}, we obtain the dependence of the tensor-to-scalar ratio, $r$, and the spectral 
tilt, $n_s$, as a function of $Q$ at Hubble exit. We again consider  both cases of nonthermalized and
thermalized inflaton perturbations. The corresponding results are presented in {}Fig.~\ref{fig4}.
One notices from {}Fig.~\ref{fig4}(c) that our results agree with the ones shown in the {}Fig. 1 of Ref.~\cite{Berghaus:2025dqi} in the region around $Q_* \in [0.2,15]$. But they deviate significantly for $Q\lesssim 0.2$ and $Q_* \gtrsim 15$.
The difference in the results for the weak dissipative regime can be attributed due to the unsuitability
of the fitting function for the power spectrum considered in Ref.~\cite{Berghaus:2025dqi} for that region of $Q$
values, while the difference
at larger values of $Q_*$ can be attributed to the fact of the inflaton field amplitude dependence in the dissipation coefficient Eq.~(\ref{Upseff}) becoming more prominent
in the large dissipative regime of WI. 
This change of behavior in the spectrum at large $Q_*$ turns out to happen
here earlier than the estimate of $Q_*> 40$ quoted in Ref.~\cite{Berghaus:2025dqi}. 
In fact, as $Q$ increases, the leading temperature dependence power
in $\Upsilon_{\rm eff}$ shifts to values lower than three, $\Upsilon_{\rm eff} \propto T^c$, with
$c<3$. The power spectrum in this case is expected to have less power than in the case of $c=3$~\cite{Graham:2009bf,Bastero-Gil:2011rva}, forcing $n_s$ to become redder. 

The range of values for $Q_*$ which falls inside the observational window are $Q_*\in [0.0076, 30]$.
{}For $Q_* < 0.0076$ the tensor-to-scalar ratio becomes larger than the upper bound set by the
BICEP, Keck Array and Planck combined data~\cite{BICEP:2021xfz}, $r<0.036$, while for $Q_* \gtrsim 30$
the spectral tilt goes outside of the two-sigma region for the same combined data.

As a note, we point out that the recent released data results from the Atacama Cosmology Telescope 
(ACT)~\cite{ACT:2025tim} push the results for $n_s$ to slightly larger values compared to the
earlier results from the Planck/BICEP and Keck ones.
While in the WI models studied in Ref.~\cite{Berera:2025vsu} those update results tend to favor
slight higher values of $Q$, in the present model they go in the opposite direction, as seen here in the
{}Fig.~\ref{fig4}(c).
 
{}Finally, in {}Fig.~\ref{fig5}, we present the results for some of the background quantities of
interest as a function of $Q_*$, namely, the temperature over the Hubble parameter, $T/H$, the Hubble parameter, 
the temperature and the axion decay constant $f_a$. The value of $f_a$ is obtained from the 
coefficient of the effective dissipation coefficient
and  defined through Eq.~(\ref{CUps}).
One notices in particular that in the observational window of valid $r$ and $n_s$ values, we have 
 $f_a \in[4.98 \times 10^{10}, 1.23\times 10^{13}]$GeV, which is still consistent with 
present bonds on the QCD axion for instance~\cite{Marsh:2015xka,DiLuzio:2020wdo}.

%%%%%%%%%%%%%%%%%%%%%%%%%%%%%%%%%%%%%%%%%%
\section{Conclusion}  
\label{conclusions}

In this paper, we revisit the proposal of Ref.~\cite{Berghaus:2025dqi} to construct a WI scenario from
SM gauge interactions. The embedding of WI within the SM framework represents a significant 
advancement, as it establishes, for the first time, a concrete connection between WI dynamics 
and the particle content of the SM. We reanalyze the results of Ref.~\cite{Berghaus:2025dqi} 
using the \texttt{WI2easy} code, a numerical tool designed for precision computation of background evolution 
and perturbations in WI. {}Furthermore, we critically reassess key assumptions in the original work. 
Specifically, we scrutinize the validity of the approximate power spectrum employed in that reference by 
comparing it to results obtained with \texttt{WI2easy}, incorporating the full temperature and 
inflaton amplitude dependencies of the dissipation coefficient. Additionally, we evaluate whether 
inflaton perturbations could undergo thermalization --- a crucial consideration for ensuring the accuracy 
of WI power spectrum predictions and the reliability of derived observational parameters. 
These refinements address pivotal challenges in achieving robust, first-principles calculations in WI scenarios.

Our results indicate that the inflaton can achieve a thermal equilibrium distribution for values
of the dissipation ratio at Hubble exit of $Q_* \gtrsim 0.08$, hence already in the weak dissipative regime
of WI. While our more accurate results obtained with  \texttt{WI2easy} show agreement with the
results from Ref.~\cite{Berghaus:2025dqi} for values of $Q_*$ between approximately 0.6 and 15, there are
significant differences in the $Q_* \lesssim 0.6$ and $Q_* \gtrsim 15$ regions. We have identified
the source of these differences, which can be attributed to the approximations considered in Ref.~\cite{Berghaus:2025dqi}. 
 
{}For future work, it would be of interest to also analyze possible non-Gaussianity
signatures that this realization of WI might have, following for example the methods set forward in Ref.~\cite{Bastero-Gil:2014raa}. 
This would need a more dedicated analysis using the form of the dissipation
coefficient given by Eq.~(\ref{Upseff}), which differs from the most simpler form  where $\Upsilon \propto T^c$.
If one assumes the case with $c=3$ analyzed in Ref.~\cite{Bastero-Gil:2014raa}, for the range of  $Q_*$  values of interest that we have obtained here,
$Q_*\in [0.0076, 30]$, the estimated non-Gaussianity coefficient for the characteristic shape of WI obtained in Ref.~\cite{Bastero-Gil:2014raa} is
$5 \lesssim |f_{\rm NL}^{\rm warm}| \lesssim 30$ (for a
more recent analysis for this case, see also Ref.~\cite{Mirbabayi:2022cbt}). 
We expect that Eq.~(\ref{Upseff}),
which has an effective power in the temperature $c< 3$, to produce a smaller $ |f_{\rm NL}^{\rm warm}| $.
Observational constraints based on the
warm shape of the bispectrum in WI are, unfortunately, very poorly known at the moment, with results derived only for the 
case of the bispectrum obtained in the very simple case
of a constant dissipation coefficient (i.e., with no temperature dependence on $\Upsilon$), with the result
for $f_{\rm NL}^{\rm warm} $ estimated to be~\cite{Planck:2019kim} $|f_{\rm NL}^{\rm warm}| = 48 \pm 27$.

In conclusion, our findings establish the viability of constructing successful WI scenarios directly from
SM gauge interactions. {}Furthermore, they highlight the critical role of the \texttt{WI2easy} numerical 
toolkit in achieving reliable results: while approximate analytical methods remain valuable 
for conceptual studies, our work demonstrates that bypassing such approximations is essential for deriving observationally robust predictions. This precision is particularly crucial when computing observable 
quantities---such as the primordial power spectrum or spectral index---that must align with high-accuracy cosmological datasets.

%%%%%%%%%%%%%%%%%%%%%%%%%%%%%%%%%%%%%%%%%%%%%
\begin{acknowledgements}

R.O.R. acknowledges financial support by research grants from Conselho
Nacional de Desenvolvimento Cient\'{\i}fico e Tecnol\'ogico (CNPq),
Grant No. 307286/2021-5, and from Funda\c{c}\~ao Carlos Chagas Filho
de Amparo \`a Pesquisa do Estado do Rio de Janeiro (FAPERJ), Grant
No. E-26/201.150/2021. 
The work of G.S.R. is supported by a PhD scholarship from FAPERJ.

\end{acknowledgements}

\section*{DATA AVAILABILITY}

The data that support the findings of this article are openly available~\cite{data}.

%%%%%%%%%%%%%%%%%%%%%%%%%%%%%%%%%%%%%%%%%%%%%%%%%%

\end{document}